\documentclass[10pt]{iopart}
\usepackage{amsmath}
\usepackage{amsfonts}
\usepackage{dcolumn}                             
\usepackage{amssymb}
\usepackage{graphicx} 
\usepackage{dcolumn}  
\usepackage{bm,bbm}       

\usepackage{color}

\newif\ifnocolor
\nocolortrue

\newcommand{\bea}{\begin{eqnarray}}
\newcommand{\eea}{  \end{eqnarray}}
\newcommand{\bit}{\begin{itemize}}
\newcommand{\eit}{  \end{itemize}}

\newcommand{\be}{\begin{equation}}
\newcommand{\ee}{\end{equation}}

\newcommand{\brac}[1]{\langle #1|}
\newcommand{\bra}[1]{\langle #1}
\newcommand{\ket}[1]{|#1\rangle}
\newcommand{\ktimes}{\rangle\! \langle}
\newcommand{\op}[2]{|#1\ktimes #2|}
\newcommand{\opoo}{\op{0}{0}}

\newcommand{\opii}{\op{1}{1}}

\newcommand{\rhoe}{\rho_{\rm env}}
\newcommand{\env}{{\rm env}}

\newcommand{\Mp}{{\cal M}_{\rm p}}
\newcommand{\Mm}{{\cal M}_{\rm m}}

\def\bra#1{{\langle#1|}}
\def\ket#1{{|#1\rangle}}

\def\e{{\rm e}}

\def\tr{{\rm Tr}}

\def\P{{\hat P}}

\def\ODR{f_{_{\rm DR}}(t)}

\newcommand{\equa}[1]{Eq.~(\ref{#1})}

\renewcommand{\e}{\text{env}}
\newcommand{\s}{{\rm sys}}

\begin{document}
\title{Quantum non-Markovian behaviour at the chaos border}
\author{Ignacio Garc\'{\i}a-Mata$^{1,2}$, Carlos Pineda$^3$, Diego A. Wisniacki$^4$}
\address{$^1$ Instituto de Investigaciones F\'isicas de Mar del Plata (IFIMAR, CONICET), 
Universidad Nacional de Mar del Plata, Mar del Plata, Argentina.}
\address{$^2$Consejo Nacional de Investigaciones Cient\'ificas y Tecnol\'ogicas (CONICET), Argentina}                         
\address{$^3$ Instituto de F\'isica, Universidad Nacional Aut\'onoma de M\'exico, 
M\'exico D.F., 01000, M\'exico}
\address{$^4$ Departamento de F\'{\i}sica ``J. J. Giambiagi", 
             FCEN, UBA and IFIBA, CONICET, Pabell\'on I, Ciudad Universitaria, 1428 Buenos Aires, Argentina}
\ead{i.garcia-mata@conicet.gov.ar}
\date{submitted on May 27, 2013} 

\begin{abstract}
In this work we study the non-Markovian behaviour of a qubit coupled to an
environment in which the corresponding classical dynamics change from integrable
to chaotic.  We show that in the transition region, where the dynamics has both
regular islands and chaotic areas, the average 
non-Markovian behaviour is enhanced to values even larger than in the regular regime. 
This effect can be related to the non-Markovian behaviour as a function of the
the initial state of the
environment, where maxima are attained at the regions 
dividing separate areas in classical phase space, particularly at 
the borders between chaotic and regular regions. 
Moreover, we show that the fluctuations of the fidelity of the environment -- which determine the non-Markovianity measure -- give a precise image of the classical phase portrait.
\end{abstract}

\maketitle
\section{Introduction} 
The theoretical and experimental study of decoherence \cite{JoosBook,ZurekRMP} is important for -- at least -- two reasons. 
On the one hand to understand
the emergence of classicality in the quantum framework. On the other hand, to assess and 
minimize the restrictions it imposes the development in new technologies being developed related to
quantum information theory.
Decoherence appears as the result of uncontrollable (and unavoidable) interaction between a quantum system 
and its environment. The expected effect is an exponential decay of quantum interference.
Generally the theoretical approach is by means of a theory of open quantum systems \cite{BreuerBook}.
The idea is to precisely divide the total system into system-of-interest plus environment and 
then discard the environment variables and derive en effective dynamical equation for the reduced 
system state. Obtaining and solving the effective equation is generally a very difficult task, so 
approximations are usually made. 
The Born-Markov approximation, which among other things assumes weak system-environment coupling 
and  vanishing correlation times in the
environment yields a Markov -- memory-less -- process, described by a semigroup of completely positive maps. The generator of these maps is given by the 
 Lindblad-Gorini-Kossakowski-Sudarshan master
equation \cite{Lindblad,Gorini1976}. Lately, though, interest in  problems
where the Markov approximation is no longer valid have flourished
(see~\cite{BreuerBook,Daffer2004,Breuer2009,Rivas2010,Znidaric2011,Horacio2011,Haikka2012,
Clos2012,Alipour2012}, to name just a few).
One very interesting feature of
non-Markovian evolution is that information backflow can bring back coherence
to the physical system and sometimes even preserve it
\cite{Horacio,Zhang2013,Bylicka2013}. 

Being able to quantify
the deviation of markovianity (beyond a yes/no answer) is of importance
to compare theory and experiment, specially in circumstances where the
usual approximations (for example infinite size environment and weak coupling) 
start to break up.  
This leads to a proper 
understanding and the possibility to engineer
non-Markovian quantum open systems which has many potential applications like quantum
simulators \cite{barreiro2011}, efficient control of entanglement
\cite{Diehl2008,Krauter2011} and entanglement engineering \cite{Huelga2012}, quantum 
metrology \cite{Chin2012}, 
or dissipation driven quantum
information\cite{Verstraete2009}, and even quantum coherence in biological systems
\cite{Ishizaki2009}.  The simulation of non-Markovian dynamics and
the transition from Markovian to non-Markovian has been recently reported in
experiments~\cite{BreuerNatPhys2011,Chiuri2012,Liu2013}.

There has been much activity in this context but one basic question remains
untouched, namely that of the influence of the underlying classical dynamics of
the environment on the central system. Intuition indicates that a chaotic
environment should result in Markovian dynamics and a regular (or integrable)
environment in strong non-Markovian effects. In~\cite{garciama2012} this
transition has been studied. However, understanding the case where the
environment has associated classical dynamics consisting of a mixture of
regular islands, broken tori and hyperbolic dynamics, is still an open problem.
The importance of this case is not to be overlooked being that mixed systems
are the rule rather than the exception~\cite{ozoriobook}.  

The purpose of this work is then to shed some light on the relation between the
classical dynamics of the environment and its markovianity, for environments
where the transition from regular to chaotic is tunable by a parameter.  The
complexities of such systems make analytical treatment almost impossible so we
shall mainly focus on numerical simulations.  In order to have access to some
analytic results the central system will be the simplest possible.  We center
on a system consisting on a qubit coupled with an environment in a pure
dephasing fashion.  In such a way that the environment evolution is conditioned
by the state of the qubit
\cite{Zurek2002,Quan2006,Lemos2011,Znidaric2011,Lemos2012,Haikka2012,garciama2012}.
The qubit acts as a probe that can be used to extract important information
from the dynamics of the environment \cite{Poulin2004}. As environment we use
paradigmatic examples of quantum chaos: quantum maps on the torus. In
particular we focus kicked maps which by changing one parameter one can go from
integrable to chaotic.  

To quantify non-Markovianity we use the measure proposed in \cite{Breuer2009} which is based
on the idea of information flow, from and to the system. In our model this measure 
is directly related to the fidelity fluctuations of the environment.
The time dependence of fidelity fluctuations can be used to extract important information
about the dynamics of quantum chaotic systems, like the Lyapunov exponent \cite{Petitjean2005}.
For localized initial states the fidelity
decay and fluctuations can be extremely state-dependent \cite{Weinstein2005}.
We found that the transition from integrable
(``non-Markovian") to chaotic (``Markovian") is not uniform.  
In the
transition  there is a maximum which can be larger than the value that this
measure attains for the regular dynamics.  But more importantly, that the
maximum happens at a value of the parameter that is critical in the
corresponding classical dynamics, like the breakup of the last irrational
torus, and the onset of unbound diffusion.

We show that the non-Markovian measure used reproduces the intricate 
structure of the classical phase space
with extraordinary precision.  Moreover, we observe that the values of
non-Markovian measure as a function of position in phase space are enhanced in
the regions that are neither chaotic nor regular, i.e. at the borders between
chaos and regularity.  This establishes the non-Markovianity measure used,
which depends on the  long time fidelity fluctuations, as pointer to the chaos
border. Another way of identifying this border can be found in
\cite{Weinstein2002border}.  As a consequence, our results contribute to a
deeper understanding of the fidelity decay  of quantum systems with mixed
classical dynamics, which is an open problem of current interest
\cite{Gorin2006,Jacquod2009,DiegoScholar}. 

This paper is organized as follows. In Sec. II we introduce the definition and
the measure of non-Markovianity that we use throughout the paper.  Then in Sect
III we describe the way that our model environment interacts with the central
system which is a qubit. We explicitly write the dynamical map and show how the
non-Markovianity measure we chose to use depends on the fidelity of the
environment. In Sec. IV we give a brief description of the quantum maps that
we use a model environments. Depending on a parameter the corresponding
classical dynamics of these maps can go from integrable to chaotic.  In
sections V and VI we show numerical results. In Sec. V for the environment in
a maximally mixed state. In Sec. VI for the environment initially in a pure
state. On average both cases show qualitatively similar results. In addition in
Sec. VI we show how the classical phase space structure is obtained when the
non-Markovianity measure is plotted as a function of the initial state. We draw
our conclusions in Sec. VII and we include an appendix where we explain some
technical details of the short time behaviour of the fidelity decay.

\section{Measuring Non-Markovianity: information flow} 
The notion  of Markovian evolution, both classical and quantum is associated
with an evolution in which memory effects are negligible. 
In classical mechanics this is well defined in terms of multiple-point
probability distributions.  In quantum mechanics evolution of an open system is
often assumed to be well described by a Lindblad
master equation (which can also be credited to Gorini, Kossakowski and Sudarshan \cite{Lindblad,Gorini1976}). The
Lindblad master equation generates a one parameter family of completely
positive, trace preserving (CPT) dynamical maps, also called quantum dynamical
semigroup.  The semigroup property implies lack of memory. But the validity of
the Lindblad master equation description relies heavily on the Born-Markov
approximation, and other restrictions.  Unfortunately there are many cases in
which these approximations do not apply, especially when weak coupling is no
longer valid, but also in the case of finite environments.  One of the key
issues is to consistently define and quantify non-Markovian behaviour for
quantum open systems.  
Recently there have been some attempts to define and quantify
non-Markovianity (some of them are reviewed in~\cite{BreuerReview}).  One of
these attempts is based on the fact that Markovian systems contract, with
respect to the distance induced by the norm-1, the probability
space~\cite{Vacchini2011}. This is often interpreted as an information leak from the system into an environment, 
as one decreases with time the ability to infer the initial condition  
from the state at a given time.  The very same idea has been used in quantum systems.  Distinguishability
between quantum states is quantified with the trace norm~\cite{Heinosaari2008},
and whenever this quantity increases with time is interpreted as 
a measure of non-Markovian beahvior 
in the quantum system by Breuer, Laine and Piilo (BLP) 
in~\cite{Breuer2009}. As this quantity is related to an information flow, and is simple in our case 
to calculate, we are going to use it to quantify non-Markovianity.

The first step is to define a way to distinguish two states. We do that by means of the trace distance. Given two arbitrary states represented by their density matrices $\rho_1$ and $\rho_2$ the trace distance is defined by
\begin{equation}
D(\rho_1,\rho_2)=\frac{1}{2}\tr |\rho_1-\rho_2|.
\end{equation}
It is a well defined distance measure with all the desired properties and it can be shown to be a good measure of distinguishability \cite{HayashiBook}. Another property of the trace distance is that 
\begin{equation}
D(U\rho_1U^\dagger,U\rho_2U^\dagger)= D(\rho_1,\rho_2)
\end{equation} 
i.e. it is invariant under unitary transformations and is a contraction
\begin{equation}
D(\Lambda\rho_1,\Lambda\rho_2)\leq D(\rho_1,\rho_2)
\end{equation}
for any CPT quantum channel $\Lambda$.
Thus no CPT quantum  operation can increase distinguishability between quantum states. 
The idea proposed by BLP is that under Markovian dynamics the information flows
in one direction (from system to environment) and two initial states become increasingly indistinguishable. Information flowing back to the system would allow for memory effects to manifest. A process is then defined to be non-Markovian if at a certain time the distance between to states increases, or 
\begin{equation}
\sigma(t,\rho_{1.2}(0))\equiv \frac{d}{dt}D(\rho_1(t),\rho_2(t))>0.
\end{equation}
With this in mind  non-Markovian behaviour can be quantified by \cite{Breuer2009} 
\begin{equation}
\label{def_noma}
{\cal M}=\max_{\rho_{1,2}(0)}\int_{\sigma>0} dt\sigma(t,\rho_{1.2}(0)),
\end{equation} 
i.e. the measure of the total increase of distinguishability over time. The maximum is taken over all possible pairs of initial states.
 
 We should remark here that there are many other proposed measures. Rivas, Huelga and Plenio (RHP) \cite{Rivas2010} proposed two measures which are based on the evolution of entanglement to an ancilla, under trace preserving completely positive maps. There are others based on Fisher information \cite{Lu2010} or  the validity of the semigroup property \cite{Wolf2008}. 
For some situations \cite{Zeng2011} BLP and RHP are equivalent.
In our case, it is easy to see that the RHP measure, which relies on monotonous decay of entanglement in Markovian processes,  differs from the BLP measure by a constant factor.
So in the present work we only consider BLP.
\section{Non-Markovianity and fidelity fluctuations} 
\label{sec:NoMaFid}
We assume that the interaction between the environment and the probe
qubit is factorizable, and that it commutes with the internal hamiltonian of
the qubit. Neglecting the qubit Hamiltonian, by selecting the appropriate
picture, and choosing a convenient basis, one can write the Hamiltonian 
as
\begin{equation}
H= \mathbbm{I} \otimes H_\env + \delta \sigma_z \otimes V.
\label{eq:H}
\end{equation}
Properly rearranged, one can write the hamiltonian of the form
\begin{equation}
H=\opoo\otimes H_0+\opii\otimes H_1,
\label{eq:H}
\end{equation}
were $H_0$ and $H_1$ act only on the environment and $\opoo$, $\opii$ are
projectors onto some orthonormal basis of the qubit  \cite{Zurek2002}. 
In this case, the coupling $\delta V= (H_0 -H_1)/2$ is given by the difference
of the hamiltonians of the environment in Eq.~(\ref{eq:H}).
This
kind of pure dephasing  interactions occur spontaneously in several experiments
(for example~\cite{NMRRMP}), but can also be engineered~\cite{Britton2005,Lemos2012}.

We suppose that initially system and environment are not correlated, which can be expressed as  
$\rho_{\s, \e}(0)=\rho_\s(0) \otimes \rho_\e$.  
To focus only on the system, the environment degrees of freedom should be traced out
\begin{equation}
\label{eq:rhot}
\rho_\s(t)=\tr_\e \left[U(t)\rho_\s (0)\otimes \rho_\e U^\dagger(t)\right]
\end{equation}
with 
\begin{equation}
U(t)=\opoo U_0(t)+\opii U_1(t).
\label{eq:Us}
\end{equation}
This yields a dynamical map for the qubit that we write as
\begin{equation}
\rho_\s(t)=\Lambda(t)(\rho_\s(0))
\end{equation}
which, in the basis of Pauli matrices, takes the form
\begin{equation}
\Lambda =\begin{pmatrix}
1 & 0 & 0 & 0 \\
0& {\rm Re}[f(t)]& {\rm Im}[f(t)]&0\\
0& {\rm Im}[f(t)]& {\rm Re}[f(t)] &0 \\
0 & 0 & 0 & 1
\end{pmatrix}.
\label{eq:paudepha}
\end{equation}
Here we have taken conventionally  $\{\sigma_i\} =\{ \mathbb{I},\sigma_{x},\sigma_ y,\sigma_z\}$ and
$\Lambda_{j,k} = 
	(1/2) \tr\left[\sigma_j U(t) \sigma_k \otimes \rho_\env 
	U^\dagger(t)\right]$.
In \equa{eq:paudepha} 
 $f(t)=\tr [\rho_\env U_1(t)^\dagger U_0(t)]$ is the expectation value of the echo
operator $ U_1(t)^\dagger U_0(t)$ . In this work we will assume that $H_1$ ($U_1$) is just a perturbation of $H_0$ ($U_0$).
If $\rho_\e$ is pure then $|f(t)|^2$ is the well known quantity called Loschmidt echo \cite{Jalabert2001}
-- also called fidelity -- which can be used to characterize quantum chaos \cite{Gorin2006,Jacquod2009,DiegoScholar}

In our case, where the system is one qubit, the states that maximize ${\cal M}$ in Eq.~(\ref{def_noma}) are pure orthogonal states lying at 
the equatorial plane on the Bloch sphere. \cite{Wissmann2012}. 
Here we consider two cases.
If the state of the environment is a pure state \cite{Haikka2012} $\rho_\e=\op{\psi}{\psi}$ then we get
\begin{eqnarray}
\label{eq:Mpack}
{\cal M}_{\rm p}(t)&=& 2 \int_{t=0,\dot{|f|}>0}^t d\tau \frac{d|f(\tau)|}{d\tau} \nonumber\\
&\equiv& 2 \sum_{i} [|f(t_i^{(\rm max)})|-|f(t_i^{(\rm min)})|].
\end{eqnarray}
where $|f(t)|=|\bra{\psi}U_1^\dagger(t)U_0(t)\ket{\psi}|$ is the square root of the Loschmidt echo and $t_i^{(\rm max)}>t_i^{(\rm min)}$ correspond to the  times of successive local maxima and minima of  $|f(t)|$. 
${\cal M}_p$ is the quantity considered in \cite{Haikka2012}. Throughout the paper, when the initial state is pure we will consider a coherent state centered at some point $(q,p)$ to be defined.

On the other hand, if we have no knowledge or control over the environment, then it will most likely be in a mixed state. 
If we assume it is in a maximally mixed state $\rhoe=I/N$ (with $I$ the identity in the Hilbert space of the environment) we get \cite{garciama2012}
\begin{equation}
	\label{eq:Mtrace}
{\cal M}_{\rm m}(t)=2 \sum_{i} [|\langle f(t_i^{(\rm max)})\rangle|-|\langle f(t_i^{(\rm min)})\rangle|],
\end{equation}
where $\langle f(t)\rangle$ is the average fidelity amplitude. If the average is done over a complete set of states then
\begin{equation}
\langle f(t)\rangle=\frac{1}{N}{\rm tr}[U_1^\dagger(t)U_0(t)]
\end{equation}
which depends only on the set of states being complete, but not on the kind of states. 

In the results  that  we present we model the dynamics of the environment
$U_{0,1}$ using quantum maps on the torus with a finite Hilbert space. 
Here, one can write Eq.~(\ref{eq:Us}) as 
\begin{equation}
U=
\begin{pmatrix} U_0 & 0 \\ 0 & U_0 \end{pmatrix}
\begin{pmatrix} \mathbbm{I} & 0 \\ 0 & U_0^\dagger U_1 \end{pmatrix}
\label{eq:coupling}
\end{equation}
so the coupling is provided by the echo operator $ U_0^\dagger U_1$.
In this
case, after some time the fidelity fluctuates around some constant value. This
causes a linear growth with time of $\Mp$ and $\Mm$ (the slope goes to zero
with the size of the Hilbert space). For this reason we follow the strategy  of
\cite{garciama2012} and consider $\Mm$ and  $\Mp$ up to some finite time $t$.
\section{Kicked maps} 
\label{sec:mapas}
For the numerical simulations we suppose that the dynamics of the environment is  given by a 
quantum map on the torus. Apart from the fact that these maps are the simplest paradigmatic 
examples of quantum chaotic dynamics, 
the ones we consider can be very efficiently implemented using fast Fourier transform. 
Due to periodic boundary conditions the Hilbert space is discrete and of dimension $N$. This 
defines an effective Planck constant $\hbar=1/(2 \pi N)$. Position states can be represented as 
vertical strips of width $1/N$ at positions 
$q_i=i/N$(with $i=0,\,\ldots,\, N-1$) and momentum states are obtained by discrete Fourier 
transform.
A quantum map is simply a unitary $U$ acting on an $N$ dimensional Hilbert space.  
Quantum maps can be interpreted as quantum algorithms and vice-versa.  In fact there exist 
efficient -- i.e. better 
than classical -- quantum algorithms for many of the well known quantum maps 
\cite{Gardiner1997,Schack1998,Bertrand2001,Weinstein2002,Levi2003,Bertrand2004}, making them 
interesting testbeds of quantum chaos in experiments using  quantum simulators 
(e.g. \cite{Weinstein2002,Chau2009,Lemos2012}).

Here we consider two well known maps with the characteristic properties of kicked systems, i.e. 
they can be expressed as
\begin{equation}
U=T(\hat p)V(\hat q).
\end{equation}
They also share the property that by changing one parameter (the kicking strength) they can be 
tuned to go from classical  integrable to chaotic dynamics.

The quantum (Chirikov) standard map (SM) \cite{DimaScholar} 
\begin{equation}
\label{standardU}
U_{K}^{({\rm SM})}=e^{-i \frac{{\hat p}^2}{2\hbar}}e^{-i \frac{K}{\hbar} \cos(2 \pi \hat x) }
\end{equation}
corresponds to the classical map
\begin{equation}
\begin{array}{lll}
p_{n+1}&=&p_n+\frac{K}{2\pi}\sin(2 \pi x_n)\\
x_{n+1}&=&x_n+p_{n+1}.
\end{array}
\end{equation}
Since we consider a toroidal phase space both equations are to be taken modulo 1.
For small $K$ dynamics is regular. Below a certain critical value $K_c$ the motion in momentum is 
limited by KAM curves. These are invariant curves with irrational frequency ratio (or winding number) which represent quasi-periodic motion, and they  are the most robust orbits under nonlinear perturbations\cite{Licht}. At $K_c=0.971635\ldots$ \cite{Greene1979}, the last KAM curve, with most irrational winding number, breaks. Above $K_c$ there is unbounded diffusion  in $p$. For 
very large $K$, there exist islands but the motion is essentially chaotic.

The quantum  kicked Harper map (HM)
\begin{equation}
U^{({\rm HM})}_{K_1,K_2}=e^{-i \frac{K_2}{\hbar} \cos(2 \pi \hat p)}e^{-i \frac{K_1}{\hbar} \cos(2 \pi \hat x) }
\label{harperU}
\end{equation}
is an approximation of the motion of kicked charge under the action of an external magnetic field \cite{Dana1995,ArtusoScholar}. 
Equation (\ref{harperU}) corresponds to the classical map
\begin{equation}
\begin{array}{lll}
p_{n+1}&=&p_n-K_1\sin(2 \pi x_n)\\
x_{n+1}&=&x_n+K_2\sin(2 \pi p_{n+1})
\end{array}
\end{equation}
From now on, unless stated otherwise we consider 
$K_1 = K_2 = K$." For $K < 0.11$, the dynamics described by the associated classical 
map is regular, while for $K > 0.63$ there are no remaining visible 
regular islands \cite{leboeuf1990}.

\begin{figure}[htb]
\includegraphics[width=\linewidth]{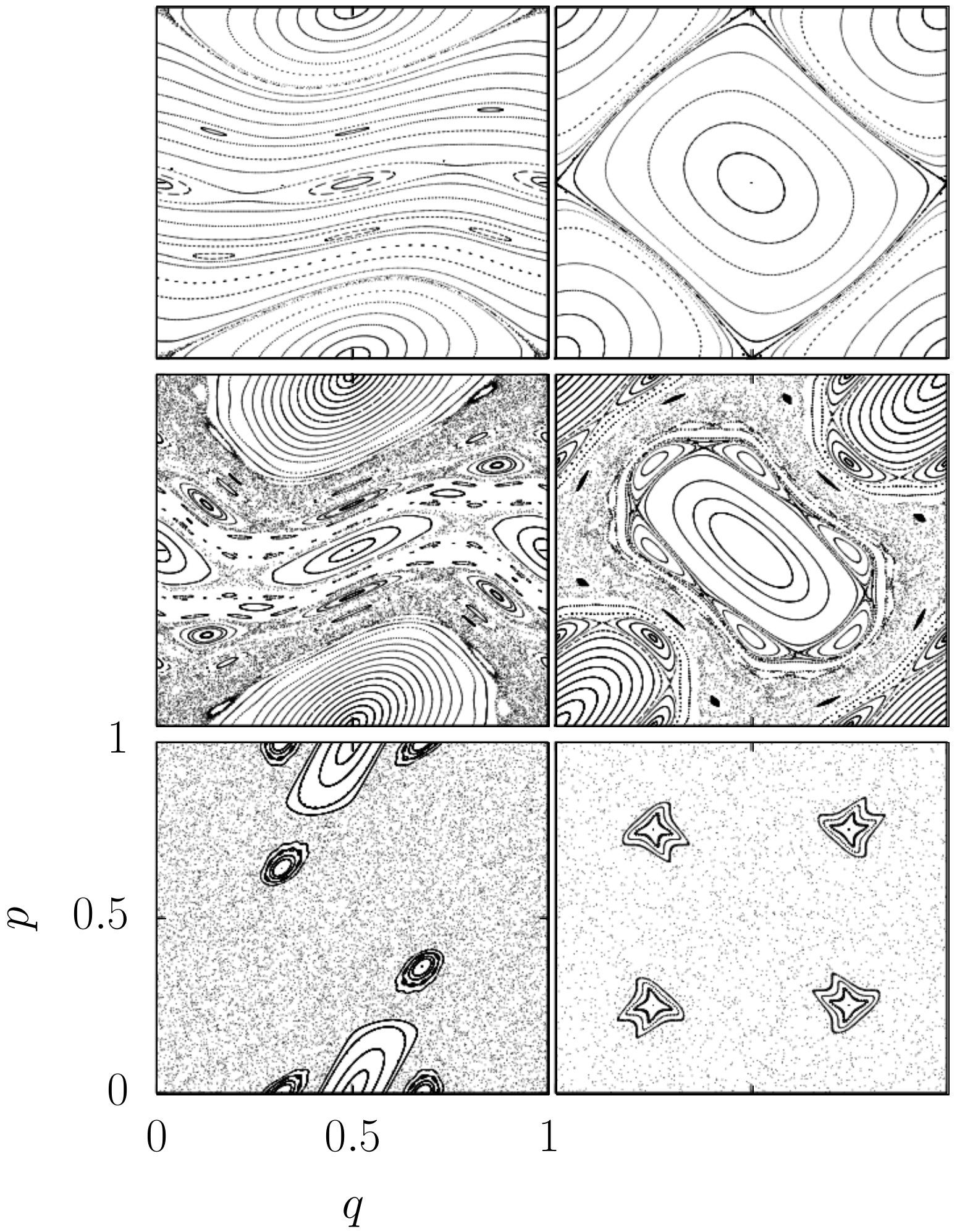} 
\caption{\label{fig:mapas} 
Phase space portrait for the classical SM (left column) and the HM, with $K_1=K_2=K$ (right column) for different values of $K$. (top, left) $K=0.5$, (top, right) $K=0.1$; (center left) $K=0.98$, (center, left) $K=0.25$; (bottom left) $K=2.5$, (bottom right) $K=0.5$ }
\end{figure}
In Fig.~\ref{fig:mapas} we show examples of phase space portraits for the two maps for three different values of $K$ where the transition from regular to mainly chaotic can be observed. 

For the numerical  calculations we take for the standard map $U_0\equiv U_{K}^{({\rm SM})}$ and  $U_1\equiv U_{K+\delta K}^{({\rm SM})}$ and for the Harper map 
$U_0\equiv U^{({\rm HM})}_{K,K}$ 
and $U_1\equiv U^{({\rm HM})}_{K,K+\delta K}$. So $\delta K$ is the perturbation strength.
\section{Non-Markovianity at the frontier between chaos and integrability}
\label{sect:Mm}
Both the SM and the HM offer the opportunity to explore the transition form integrability to chaos by changing the kicking parameter. By doing that (for the HM) two things were found in \cite{garciama2012}. 
As expected,  For very large $K$, which corresponds to chaotic dynamics, Markovian behaviour was observed. On the other hand, for small $K$ corresponding to regular dynamics, non-Markovian behaviour was obtained.
\begin{figure} 
\includegraphics[width=0.95\linewidth]{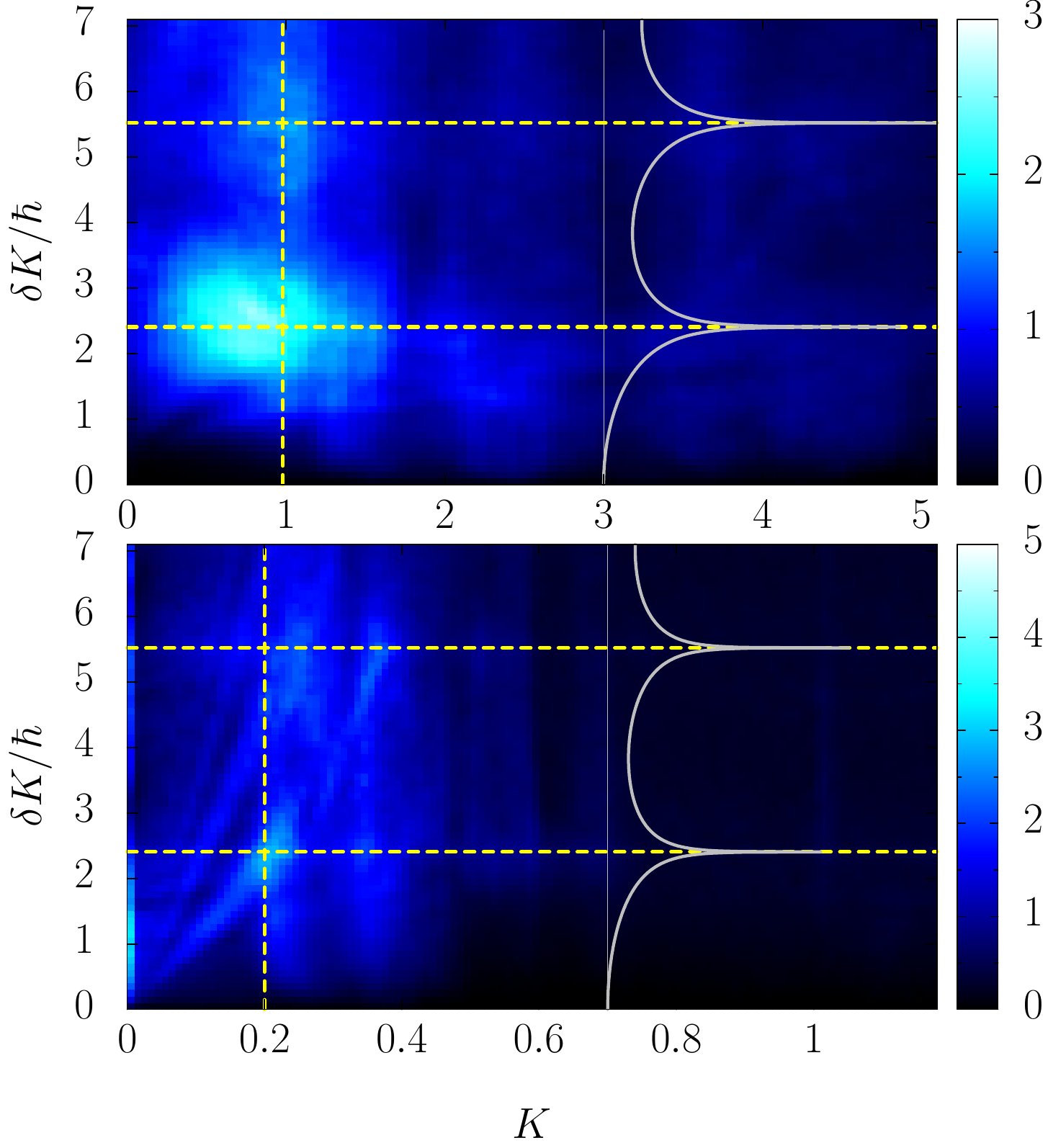}  
\caption{\label{picos2d}
${\cal M}(t=200)$ as a function of $K$ (varying the dynamics of the
      environment) and $\delta K/\hbar$ (controlling the coupling strength) for
      the quantum standard (top) and the quantum Harper map (bottom) with
      $N=500$. The vertical lines are (top) $K=0.98\approx K_c$, and (bottom)
$K=0.2$.  Horizontal dashed lines mark is the first values of $\delta K$ such
that $J_0(\delta K/\hbar)=0$. Overlay (gray/solid) curves correspond to
$\Gamma(\delta K/\hbar)$ from \equa{eq:gamma2}, rescaled to fit in the plot.}
\end{figure}
However, there was an unexpected result: the transition is not uniform. There is a clear peak in $\Mm(t)$  -- Fig. 3 in \cite{garciama2012} -- that, depending on the value of $\delta K$ and $t$ can even be  larger than the value for regular dynamics. To complement this previous result and further illustrate this effect we calculated $\Mm(t)$ as a function of $K$ and $\delta K$. 
In particular, for very short times, the decay of the  fidelity amplitude has a rich structure \cite{garciamaNJP}.
It can be shown by semiclassical calculations that for short times 
the decay of the average fidelity amplitude is given by
\begin{equation}
\label{eq:gamma}
|\langle f(t) \rangle|\sim e^{-\Gamma t}.
\end{equation}
The decay rate gamma can be computed semiclassically~\cite{garciamaNJP} and Eq.~(\ref{eq:gamma}) is valid for increasingly larger times as the system becomes more chaotic. 
For $t=1$, in the case of the HM and the SM, $\Gamma$ can be computed analytically (see appendix) and it is given by
\begin{equation}
\label{eq:gamma2}
\Gamma=-\ln |J_0(\delta K/\hbar)|,
\end{equation}
where $J_0$ is the Bessel function. Thus when $J_0(\delta K/\hbar)=0$, $\Gamma$ diverges. It can be observed that 
in fact this is the case. The fidelity amplitude decays very fast for short times, and then there is a strong revival which translates in an increase of $\Mm$ \cite{garciama2012}.

In Fig.~\ref{picos2d} we show ${\cal M}_{\rm m}(t=200)$ for the SM (top) and the Harper map (bottom).
 The horizontal axis is the kicking strength $K$ and vertical axis is the rescaled perturbation $\delta K/\hbar$.  In both cases there are clearly distinguishable maxima. The horizontal dashed lines mark the points where $\Gamma$ diverges, which is seen in the overlay plot of $\Gamma(\delta K/\hbar)$ (solid/gray lines). As expected, along those lines $\Mm$ is larger due to a large revival of the fidelity amplitude for small times.

The dashed vertical line, on the other hand, marks the position of the peak on the $K$ axis. For the SM we placed 
the line on the value $K_{\rm c}\approx 0.98$ were the transition to unbound diffusion takes place.
For the kick Harper map with $K_1=K_2$ there is no analog transition, however we see a peak near $K=0.2$.
\begin{figure}
\includegraphics[width=\linewidth]{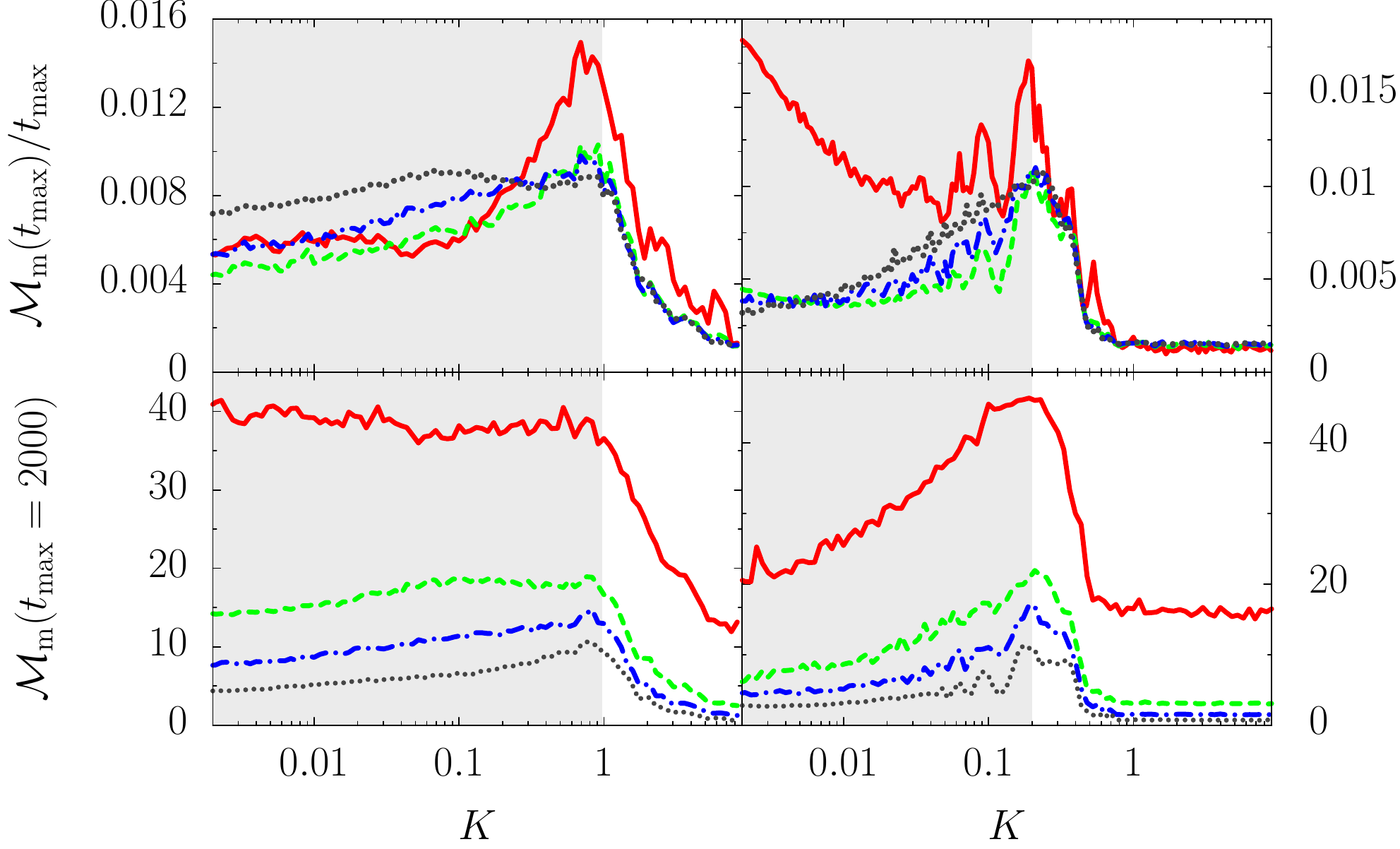} 
\caption{\label{fig:trazas}Top row: $\Mm(t_{\rm max})/t_{\rm max}$  as a function of the kicking strength $K$ with $N=512$ , for  different times $t_{\rm max}=100$(solid/red), $500$ (dash/green), $1000$ (dot-dash/blue), $4000$ (dot/gray) for
(Left) SM. (Right) HM.
Bottom row:  $\Mm(t_{\rm max}=2000)$ as a function of $K$, for different environment sizes $N=100$ (solid/red), $N=500$ (dash/green),  $N=1000$ (dot-dash/blue), and $N=2000$ (dot/gray)  for
(Left) SM and (Right) HM. In all cases $\delta K/\hbar=2.0$}
\end{figure}

In Fig~\ref{fig:trazas} we show $\Mm(t)$ as a function of $K$ for the case $\delta K/\hbar=2.0$. Panels on the left (right) correspond to the SM (HM). On the top we consider the dependence with time. It is clear that for a fixed dimension $N$, as time increases the peak establishes at a fixed value. The good scaling with $t_{\rm max}$ (after the peak) can be explained as follows:
as the environment becomes more chaotic, the fidelity decays faster and fluctuates around a constant value. As a consequence,  the growth of $\Mm(t)$ becomes linear in time -- much sooner for a chaotic environment --, with a slope proportional to ($1/N$) \cite{garciama2012}. So for a fixed $N$, the curves should scale with time. The discrepancy for short $t$ is understood because the linear regime is not yet attained.
On the bottom row of Fig~\ref{fig:trazas} we explore the possibility of finite size effects. We show $\Mm(t=2000)$ with a fixed $t$. As $N$ grows the peak settles at a constant value -- again $K\approx 0.98$ for the SM and $K\approx 0.2$ for the HM  (both marked by the limit of the shaded region).

We conclude this section by stating that $\Mm(t)$, which  depends on the fluctuations of the average fidelity amplitude, seems to be reinforced at a classically significant point in parameter space for the SM, namely $K_c$. On the other hand for the HM we make a complementary remark.
Since the peak at $K=0.2$ seems to be robust (both changing $N$ and $t_{\rm max}$, we conjecture that in analogy to $K_c$ of the SM there should be a similar transition, at least in some global property of the classical map, near $K=0.2$. We postpone that discussion until Sec.~\ref{sect:class}.
\begin{figure}
\includegraphics[width=\linewidth]{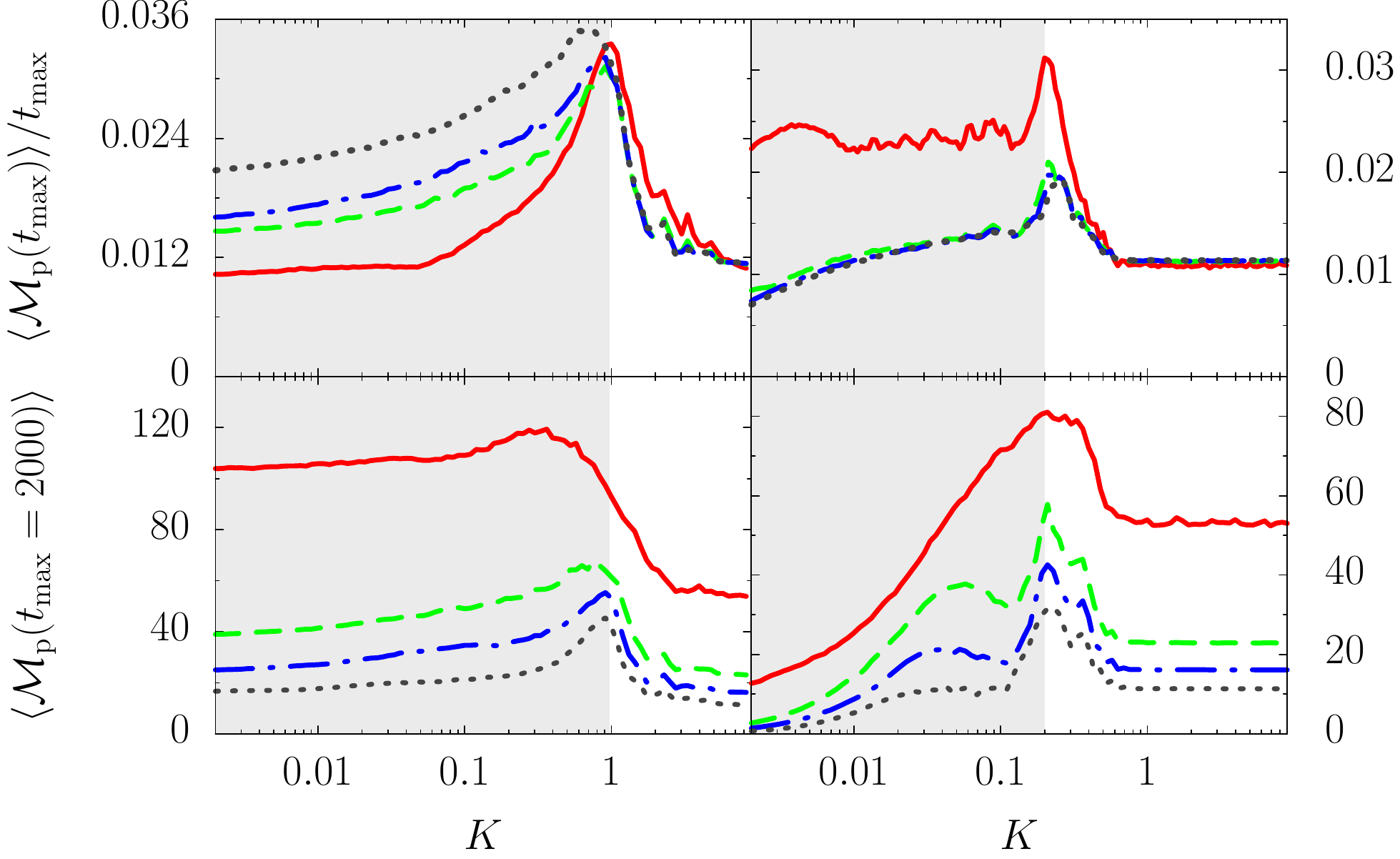} 
\caption{\label{fig:paquetes} Same as Fig.~\ref{fig:trazas} for $\Mp(t)$}
\end{figure}
\section{Environment in a pure state: classical phase space revealed} 
In the previous section we obtained unintuitive results for the non-Markovianity when the dynamics of the environment goes from integrable to chaotic. 
In particular there appears to be maxima of $\Mm$ as a function of $K$ (and $\delta K$).
To obtain these results,
we chose  the initial state of the environment to be in a maximally mixed state so 
the measure $\Mm(t)$ depends on the average fidelity amplitude (see \equa{eq:Mtrace}). This average is a sum of amplitudes and interference effects could be argued to be at the origin of the peaks observed.  
For completeness, in this section we suppose 
the environment to be initially in a pure state \cite{Haikka2012}, in particular a Gaussian -- or coherent -- state, using $\Mp(t)$ of \equa{eq:Mpack},  for two reasons. First to contrast the global properties obtained with $\Mm(t)$ , through the average behaviour of the fidelity. But also, 
to show that $\Mp(t)$, and as a consequence fidelity fluctuations, as a function of the center of the initial 
Gaussian wave packet, gives a precise image of the classical phase portrait.
\begin{figure*}
\includegraphics[width=0.85\linewidth]{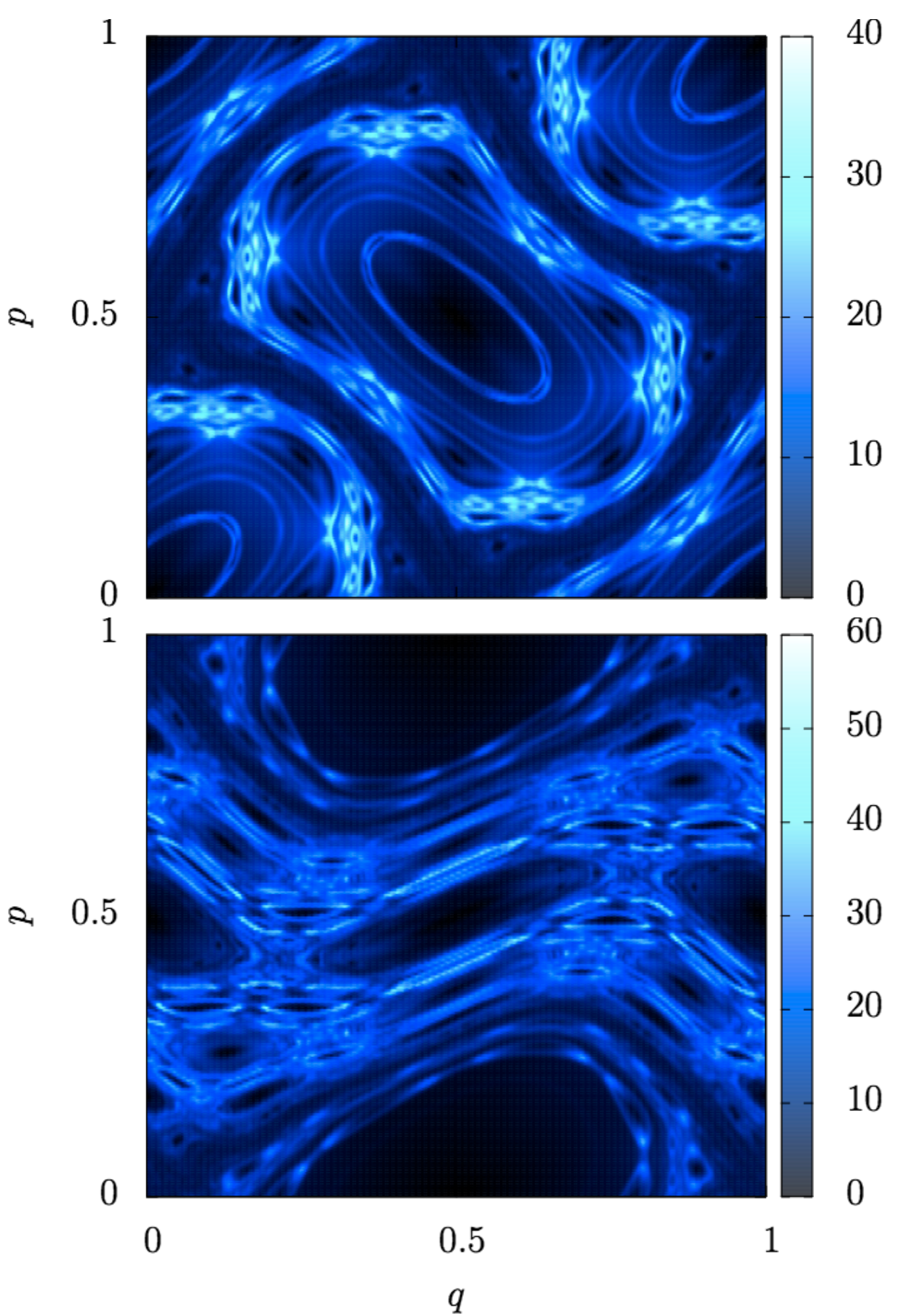} 
\caption{\label{edf_2}$\Mp(t)$ as a function of  the center $(q_0,p_0)$ of the initial Gaussian wave packet with $N=5000$, $\delta K/\hbar=2$ and $t=500$ for the SM with $K=0.98$ (bottom) and the HM with  $K=0.25$ (top).}
\end{figure*}
\subsection{Correspondence between $\Mm$ and $\langle \Mp\rangle$}
In this section 
we contrast the results for $\Mm(t)$ in  Sec.~\ref{sect:Mm} with the ones for the average of $\Mp(t)$. 
We consider a uniform grid of $N$ points at positions $q_i=i/\sqrt{N},\, p_j=j/\sqrt{N}$ and place 
coherent state centered at each pair $q_i,p_j$. 
We then average over all the initial states of the environment and get
\begin{equation}
\label{eq:avMp}
\langle \Mp(t)\rangle=\frac{1}{N}\sum_{i=0}^{\sqrt{N}-1}\sum_{j=0}^{\sqrt{N}-1}\Mp^{(q_i,p_j)}(t),
\end{equation}
where $\Mp^{(q_i,p_j)}(t)$ is just $\Mp(t)$ for a particular Gaussian state centered at $(q_i,p_j)$. 

In Fig.~\ref{fig:paquetes} we show for $\langle \Mp(t)\rangle$ the curves for the parameters 
that correspond to the ones obtained in 
Fig.~\ref{fig:trazas}. As expected, the curves are different, but  the qualitative properties are very 
similar. Mainly the marked peak at $K\approx 0.98$ for the SM and at $K\approx 0.2$ for the HM are preserved.
On the top row, we observe that after the peak the scaling with $t_{\rm max}$ also holds.
On the bottom row the dependence with $N$ is shown. It is also clear that the peak becomes more defined as $N$ grows.
 \subsection{Classical phase space sampling using $\Mp$}
We have shown (Figs.~\ref{fig:trazas} and \ref{fig:paquetes}) that qualitatively there is no 
difference between the non-Markovianity for the environment 
in a completely mixed state and the average non-Markovianity
for the environment in a pure coherent state. This is what we called global feature,
``global'' referring to an average over states covering the whole phase space.
The only difference is whether we do the average over amplitudes ($\Mm$) or
probabilities ($\Mp$) .  Nevertheless for individual pure states of the
environment, non-Markovianity is strongly state-dependent.  In this section we
seek to show this dependence is strongly related to  the details of the
classical phase space portrait of  the environment.  For this, we again define
a grid of $n_s$ points
$q_i,\,p_j=0,1/\sqrt{n_s},\ldots,(\sqrt{n_s}-1)/\sqrt{n_s}$. We then compute
$\Mp(t)$ for a fixed time, and we plot this as a function of initial position
and momentum.  In other words we plot $\Mp^{(q_i,p_j)}(t)$ from \equa{eq:avMp}.
The results obtained were surprising. We did expect that the classical dynamics
should and would have some kind of effect. However we did not expect that $\Mp$
would reproduce with such detail the complexities of the classical phase space.
In   Fig.~\ref{edf_2} it is shown how all the classical structures are very
well reproduced by the landscape built from ${\cal M}_{\rm p}(t)$ as a function
of $q_i$ and $p_j$ (see the corresponding classical cases in
Fig.~\ref{fig:mapas}).  Of course, the ability to resolve classical structure
will be limited by two factors: the dimension $N$ (or equivalently, size of
effective $\hbar$), and the number of initial states $n_s$ (or ``pixels''). In
Fig.~\ref{edf_2} it could be argued that $N=5000$ is almost classical. This
argument becomes relative when one considers that a quantum map with $N\approx
5000$ could be implemented in a quantum computer of the order of, a little more
than, 12 qubits (not a very big number of particles).

\begin{figure}
\includegraphics[width=0.9\linewidth]{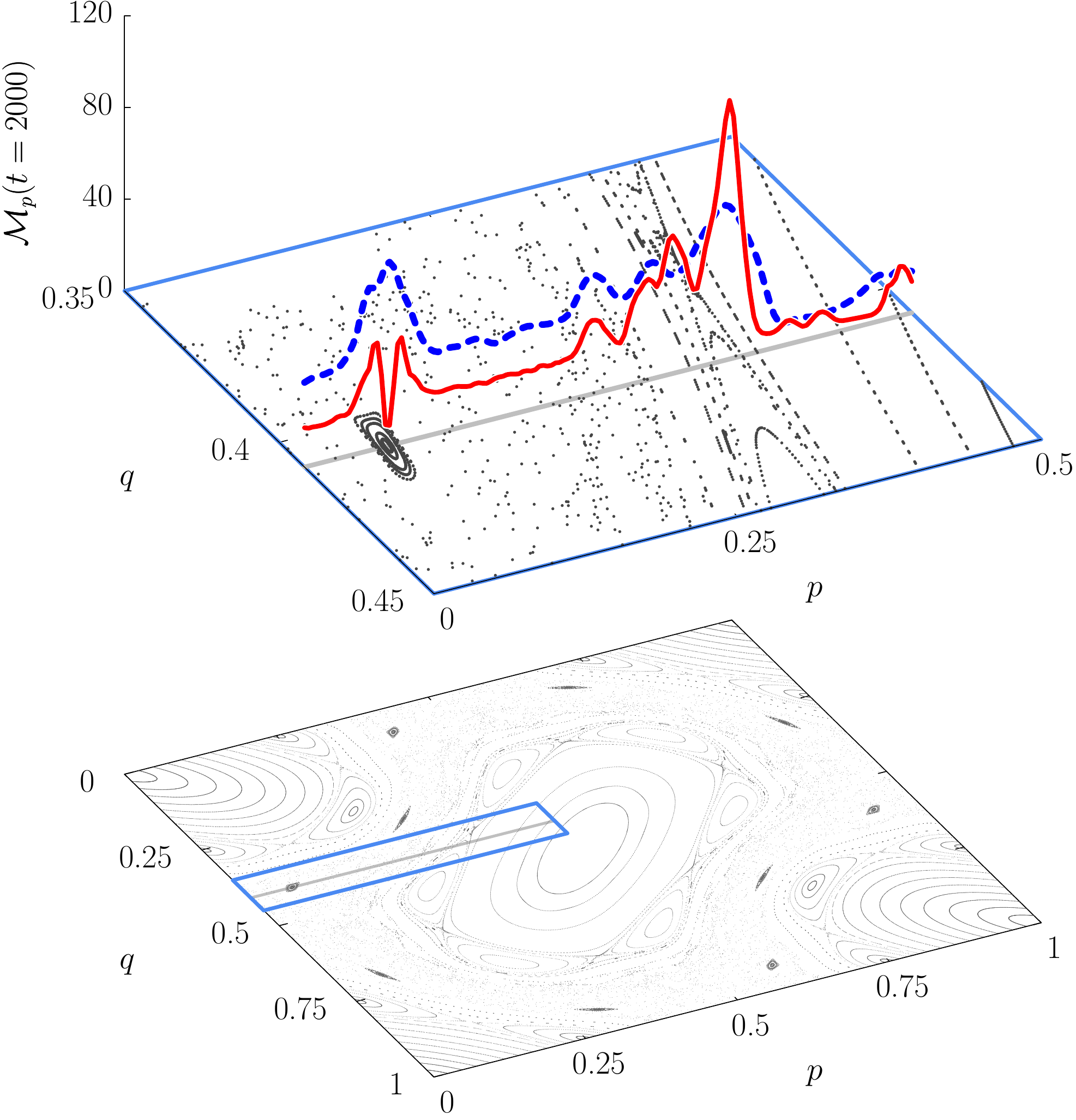} 
\caption{\label{detalle_edf}$\Mp(t=2000)$ as a function of initial position
($q_0=0.408171$, $p_0\in [0,0.5]$) of the state of the environment and two
different dimensions (red/solid) $N=4000$, (blue/dashed) $N=1000$. The map is
the quantum HM with $K=0.25$, $\delta K/\hbar=2$. The blue square on the bottom
indicates the area that is detailed above.} \end{figure}

But the most surprising thing is that, contrary to  intuition, ${\cal M}_{\rm
p}(t)$ is almost as small for a regular environment (i.e. when the initial
state is localized inside a regular island) as for chaotic a chaotic
environment. On the contrary ${\cal M}_{\rm p}(t)$ exhibits peaks at the
regions that separate different types of dynamics. Specifically at the complex
areas consisting of broken tori that separate regular islands and chaotic
regions and near hyperbolic periodic points. This means that the main
contribution to the average non-Markovianity (in the mixed phase space case)
does not come form the regular parts.  In Fig.~\ref{detalle_edf} we show a two
dimensional curve that corresponds to a detail of the HM case in
Fig.~\ref{edf_2}. Two things can be directly observed. The first one is how
$\Mp$ becomes larger and has maxima in the regions  that lie between regular
and chaotic behavior. But also how larger $N$ resolve better the small
structures. In particular elliptic periodic points are expected to be a minimum
of $\Mp$ because they are structurally stable and so a small perturbation
leaves them  unchanged. In that case fidelity does not decay, or decays very
slowly. The dashed (blue) line ($N=1000$) detects the change between regular
and chaotic, but does not resolve the structure inside the island. In this case
the width of a coherent state ($1/\sqrt{N}$) is of the order, or larger, than
the size of the island.  On the contrary, for larger $N$ (red line) the
structure inside the island is well resolved and $\Mp$ has maxima on the
borders of the island and is minimal on top of the periodic point.
It is worth pointing out, that non-Markoviantiy in the regular and chaotic
regions have very different scaling with $N$. For a fixed, large time -- $t
\sim 2000$ -- we have observed that in the chaotic region $\Mp$ decays as
$1/\sqrt{N}$, while in the regular region $\Mp$ grows as $\sqrt{N}$ (we have
observed this numerically for sizes up to $N=6\times 10^5$). In the border
regions, the behaviour has no clear scaling with $N$, as the small classical
structures are better resolved. 

A further comment on the finite-size scaling: it is known that finite N in a quantum map implies 
that at some point in time recurrences occur.  
Build-ups in fidelity can be observed at around
Heisenberg time ($T_{\rm H}\approx N$)\cite{Pineda2006}.  The contribution of
these isolated revivals are negligible as opposed to
the very frequent revivals that occur for systems in the
border regions (see Fig.~\ref{compfid_M3}). In fact, in the semiclassical limit
($N\to \infty$) these Heisenberg time revival looses all meaning.
\begin{figure}
\includegraphics[width=0.9\linewidth]{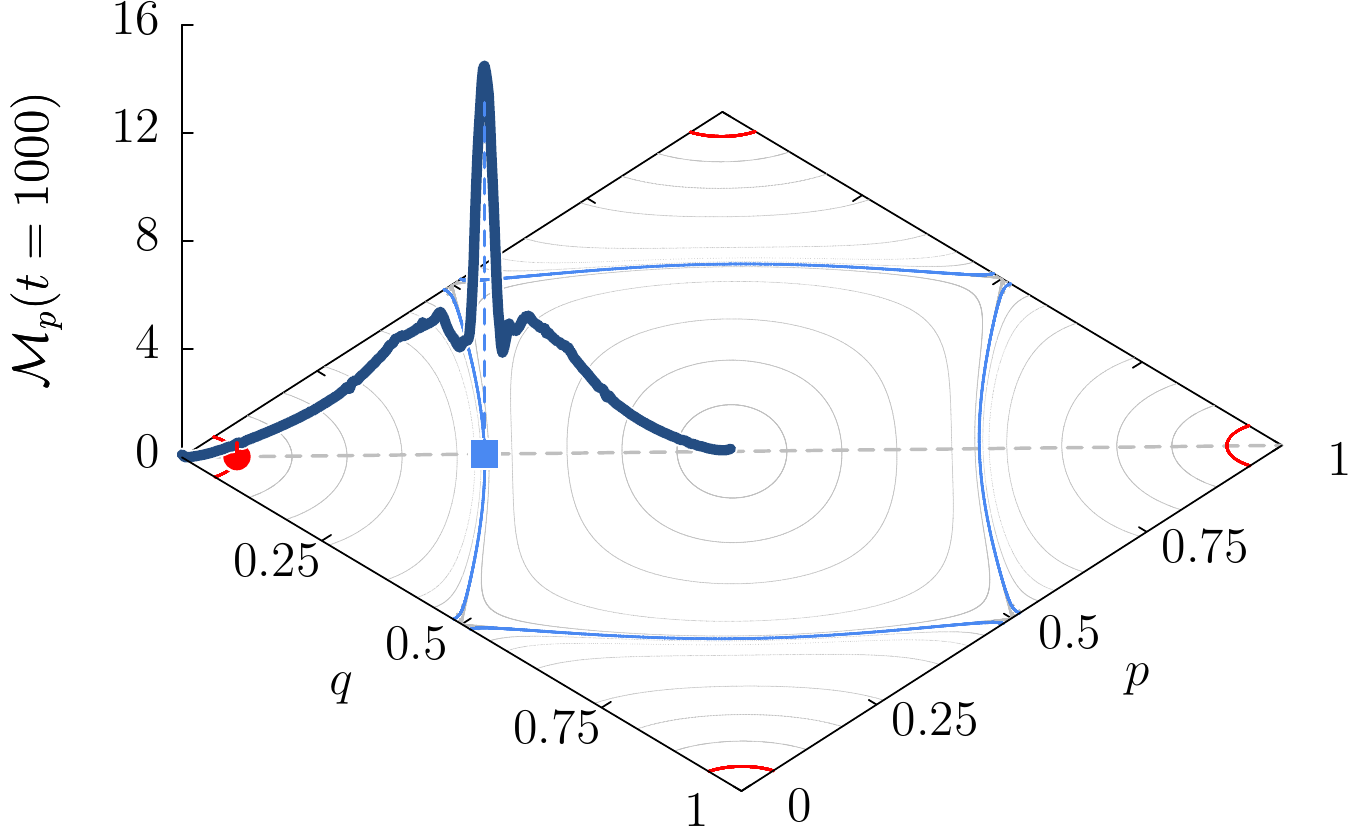} 
\caption{\label{k01QeqP} $\Mp(t=1000)$ as a function of the position $q_0=p_0$ of the initial state of the environment. The map is the quantum HM with $N=2000$, $K=0.1$, $\delta K/\hbar=2$. Dark lines are the classical trajectories corresponding to the to curves shown in Fig.~\ref{compfid_M3}.}
\end{figure}

From the numerical results we conclude that the main contribution to
non-Markovian behaviour comes from the regions of phase space that delimit two
separate regions -- like  chaotic and regular regions, and also two disjoint
regular islands.  The relation with the fidelity is key to understanding this
effect.  For states in the chaotic region the fidelity decays exponentially and
saturates at a value  which depends on the size of the chaotic area (typically
proportional of order $1/N$). The main 
contribution to non-Markovianity for chaotic initial conditions come from small
time revivals (see e.g. \cite{garciamaNJP}). The contribution due to
fluctuations around the saturation value grows linearly with time, but with a
slope that is inversely proportional to $N$, so in the large $N$ limit it can
be neglected. Gaussian initial states inside regular islands evolve in time
with very small deformation, so the fidelity is expected to decay very slowly
and eventually there will be very large (close to 1) revivals. However the
large revivals will be sparse and their contribution to non-Markovianity will
be small.  In the border areas there is no exponential spreading so the initial
decay is be slower, and there is no chaotic region so there is no expected
saturation. As a result after a short time decay we observe numerically that
there are high frequency fluctuations that contribute strongly to the non-
Markovianity measure.

We illustrate this in Figs.  \ref{k01QeqP} and \ref{compfid_M3} .  In
Fig.~\ref{k01QeqP} we show $\Mp(t=1000)$ for the HM with coherent states
centered at $q=p$.  There is a minimum at the fixed point which is understood
again in terms of structural stability: when the map is perturbed, the fixed
point remains a fixed point, so fidelity does not decay.  To further understand
we use Fig. ~\ref{compfid_M3} where we show the fidelity and $\Mp(t)$ for the
two points marked in Fig.~ \ref{k01QeqP} (circle: $q_0=p_0=0.05$; square:
$q_0=p_0=0.275$).  Inside the islands the motion of the wave packets is more or
less classical with little stretching over long times. Fidelity decays slowly
and since there is practically no deformation, there are large revivals at
times of the order of $N$ (solid/red curve in Fig.~\ref{compfid_M3} top). On
the contrary at the separatrix the  wave packets spread and though initially
fidelity decays fast, there remains a significant overlap the whole time. The
complex motion accounts for the fluctuations (dashed/blue curve in
Fig.~\ref{compfid_M3} top) and the resulting maximum of  $\Mp$ seen in Fig
.\ref{k01QeqP}.

\begin{figure}
\includegraphics[width=0.95\linewidth]{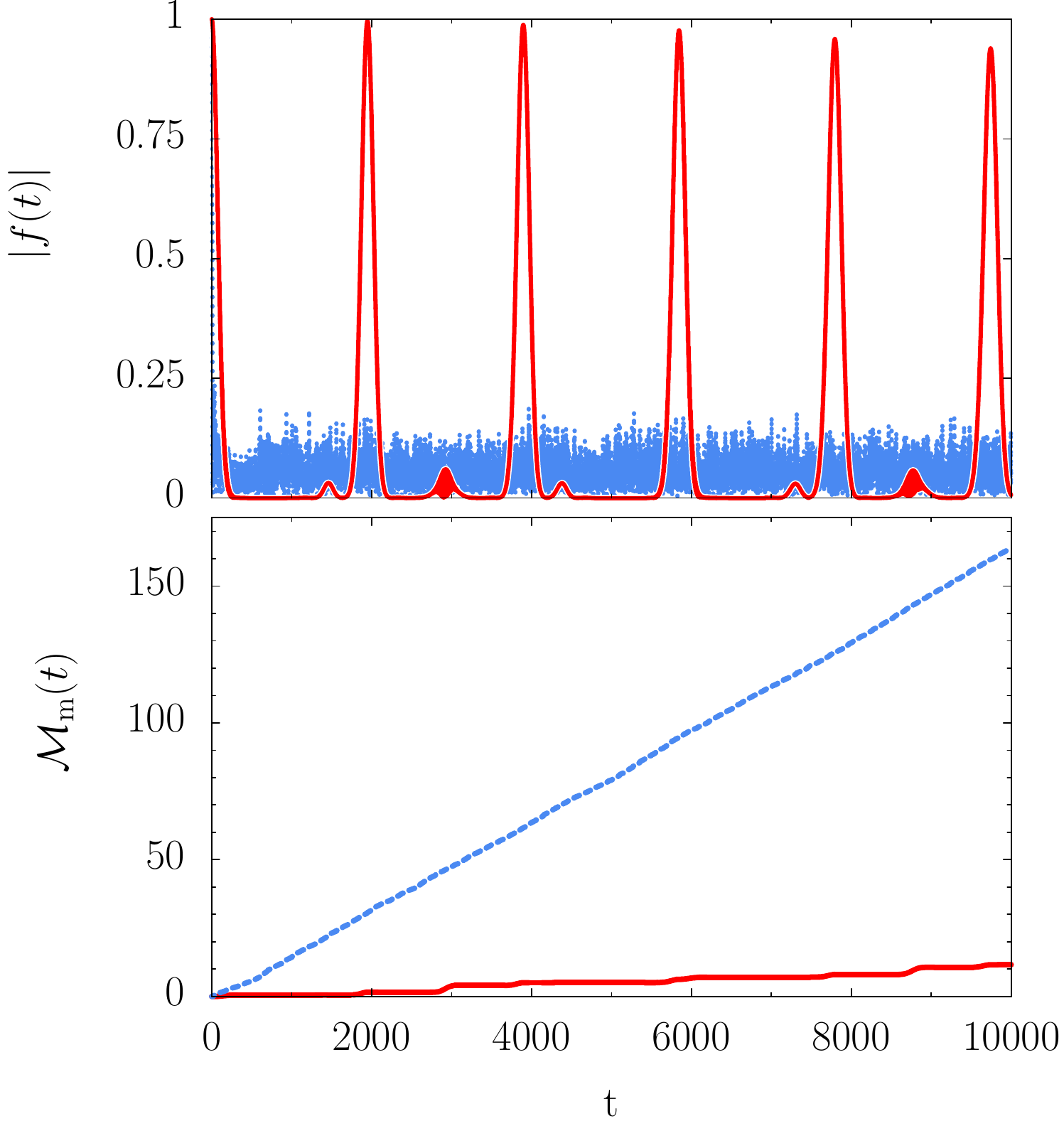} 
\caption{\label{compfid_M3} Top: $|f(t)|$ for  two coherent states evolved with the quantum HM for $K=0.1$, $\delta K/\hbar=2.0$ and $N=2000$. (blue) $q_0=p_0=0.275$ corresponds to the maximum seen in Fig.~\ref{k01QeqP}, and  (red) $q_0=p_0=0.05$.}
\end{figure}
A deeper understanding of the behaviour of the non Markovianity measure ${\cal M}(t)$ at 
the border between the chaotic and integrable region would be desirable. 
We have observed that in that region, for long times, the wave function is trapped within 
a relatively small portion of phase space. We could infer that,
for these times, there is going to be a smaller area of phase space available, 
and the wave function will behave approximately randomly with time. 
The smaller phase space available, translates into a
smaller effective Hilbert space, where the fluctuations will thus be larger.
To support this reasoning we have observed that both the Husimi and Wigner 
(without ghost images \cite{Arguelles2005})  functions of states
contributing largely to the measure of non-Markovianity remain localized inside
the sticky area near the KAM region~\cite{ding1990,karney1983}. We also tested
if the Fourier transformation of the fidelity amplitude  
is compatible with random data (i.e. has little or no structure). 
Moreover, we looked at the distribution of fidelity amplitude, which 
if it had Gaussian distribution it would be compatible with the inner 
product of two random state. However, although we have found that the distribution of the 
fluctuations of the fidelity amplitude is indeed Gaussian for many cases, the Fourier 
transform at the border exhibits some clear peaks meaning that the behaviour is not completely random, so 
it is
not simply a matter of smaller effective Hilbert space dimension. 
Work in this direction is in progress.

\subsection{Compatibility with classical results}
\label{sect:class}
The results obtained in the previous sections relate NM with some global property of the classical 
system. For the standard map there is a critical value $K_{\rm c}$ of the kicking strength after 
which the motion in the momentum direction (when the map is taken in the cylinder) becomes 
unbounded. This value, estimated to be
$K_{\rm c}=0.971635\ldots$ \cite{Greene1979},
corresponds to the breaking of the torus with most irrational winding number. 
 
Motion for the kicked Harper map  on the contrary is different. For $K=0$ it is integrable with a 
separatrix joining 
the unstable fixed points $(0,1/2),\, (1/2,0),\, (1,1/2),\, (1/2,1)$. For $K>0$ the separatrix 
breaks and -- if considered on the whole plane -- a mesh of finite width forms, also called 
stochastic web. Motion inside this mesh is chaotic and diffusion is unbounded for all $K$. 
 
Although --to our knowledge -- for the HM there is no critical $K$  analog to $K_{\rm c}$ for the 
standard map,  the peaks in Figs.~\ref{picos2d}, \ref{fig:trazas}, and \ref{fig:paquetes} hint that 
there could exist a similar kind of transition function of $K$. To test this conjecture we evaluate 
two different global quantities.  First we take into account diffusion. Considering the map on the 
whole plane (i.e. no periodic boundaries) if diffusion is normal then the spreading, e.g. in 
momentum, should grow linearly with time (number of kicks). Thus we define
 \begin{equation}
 \label{eq:diff}
 D=\lim_{t\to\infty}\frac{\langle (p_t -p_0)^2\rangle}{t},
 \end{equation} 
where the average is taken over each initial condition.
In the $t\to \infty$ limit, $D\to 0$ if diffusion is bounded. So, for the SM  we expect $D$ to be 
$0$ (or go to 0 with time) below $K_{\rm c}$ and start growing for $K>K_{\rm c}$. For the HM we do 
not know a theoretical value of $K_c$.

\begin{figure}
\includegraphics[width=\linewidth]{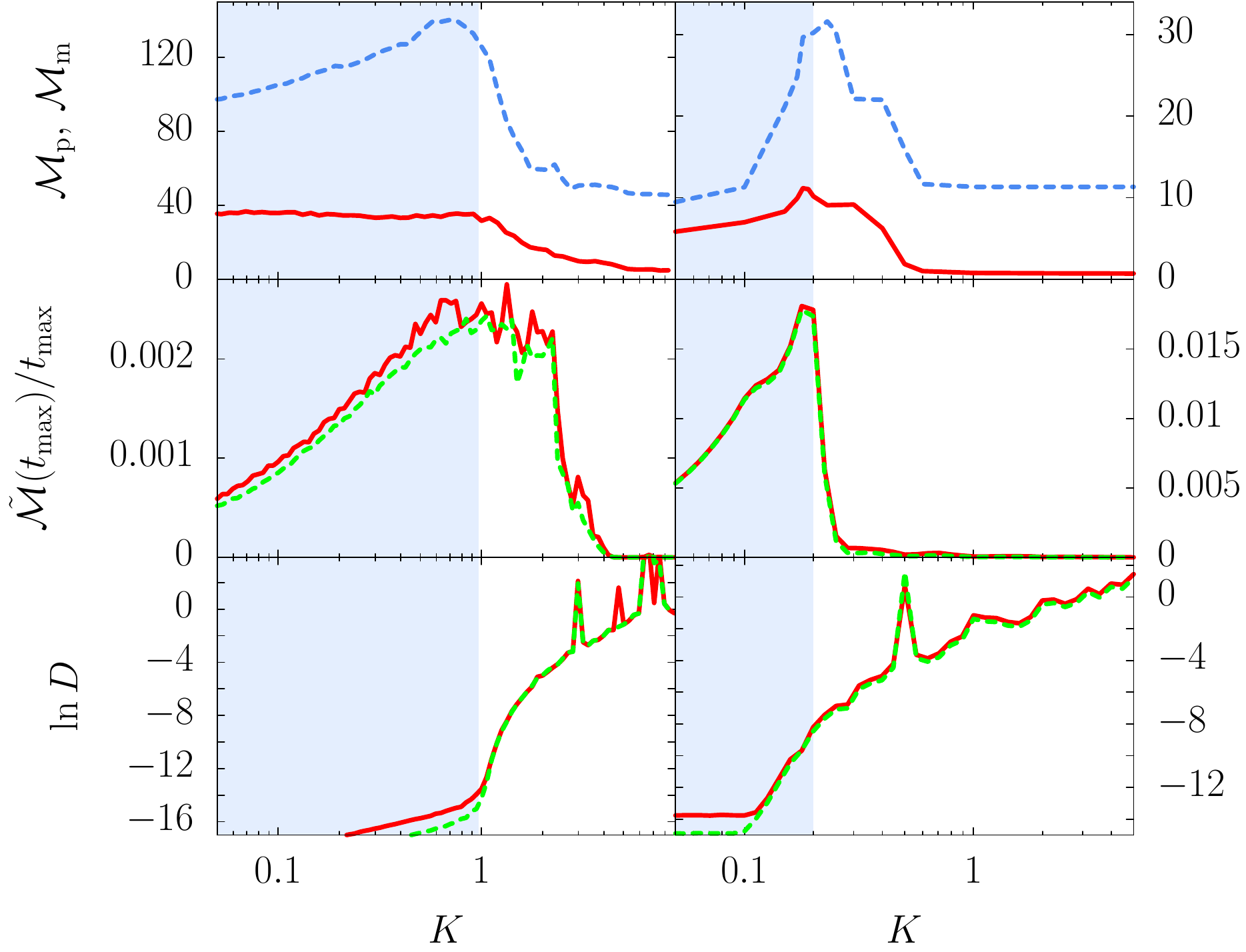} 
\caption{\label{fig:M3quant_class}Top Row: $\Mp$ (dashed/blue) and $\Mm$ (solid/red) as a function of $K$
for the SM (left) and the HM (right) with $N=512$, $\delta K/\hbar=2.0$, and $t_{\rm max}=4000$.
Middle Row: $\tilde{{\cal M}}/t_{\rm max}$ as a function of $K$ for the SM (left) and the HM (right) with $t_{\rm max}=20000$ (red/solid) and 50000 (green/dashed).
Bottom Row: $D$ as a function of $K$ for  the SM (left) and the HM (right) with $t_{\rm max}=20000$ (red/solid) and 50000 (green/dashed).
The limit of the shaded area is $k=0.98$ (SM) and $K=0.2$ HM.}
\end{figure}

However, the diffusion coefficient $D$ depends only on the unperturbed motion. We therefore propose
a measure that depends the distance between perturbed and unperturbed trajectories
and that is built to resemble ${\cal M}$.
We take an initial point $(q_0,p_0)$ an evolve it with the classical map,  without periodic 
boundary conditions, and we measure the distance 
\begin{equation}
d_t=\sqrt{(q_t-q'_t)^2+(p_t-p'_t)^2}
\end{equation}
as a function of (discrete) time  $t$, with $q',\, p'$ the perturbed trajectories. Finally in order 
to mimic the behaviour of quantum fidelity we take 
\begin{equation}
\tilde{f}_t=\exp[-d_t]
\end{equation}
which is equal to 1 for $t=0$ and decays for $t>0$. For chaotic motion, e.g. on the stochastic web 
defined by the  HM  $f_t\to 0$ as $t\to\infty$. In analogy with (\ref{eq:Mpack}) and 
(\ref{eq:Mtrace}), we 
define
\begin{equation}
\tilde{{\cal M}}(t)=\sum_{f_t-f_{t-1}>0}(f_t-f_{t-1}).
\end{equation}
The value of this quantity becomes apparent in the light of numerical results. In Fig.~
\ref{fig:M3quant_class} we
show ${\cal M}_{\rm m}$, ${\cal M}_{\rm p}$ (top row) and $\tilde{{\cal M}}$ (middle row) as a 
function of $K$ for both the SM and the HM. In the middle row we see  $\tilde{{\cal M}}$ for both 
maps for  two different times. There is 
a qualitatively  similar behaviour to ${\cal M}_{\rm p}$  (on the top row) where  $\tilde{{\cal M}}$ 
grows with $K$ until it reaches a peak at $K^*$, and then after that it decreases. We have already 
hinted that for the SM, this peak is reached for  $K^*\approx K_{\rm c}$, where the last irrational 
torus is broken, or when unbounded diffusion sets in. We know that $K_{\rm c}\approx 0.98$ 
\cite{Greene1979,DimaScholar}.  For the symmetric HM 
 there is both normal and anomalous diffusion, described in \cite{Leboeuf1998}, but
there  is \emph{a priori} no equivalent point to $ K_{\rm c}$ of the SM.

On the bottom row of Fig.~\ref{fig:M3quant_class} 
we show the numerical calculation of the diffusion coefficient $D$, defined in \equa{eq:diff}, by evolving a number of initial conditions up to a time t and compute the slope of $\langle (p_t -p_0)^2\rangle/t$. 
The red/solid line corresponds to $t=1000$ while the green/dashed line corresponds to $t=16000$. It is clear that after the critical point (marked by the shaded region) both curves for $D$ are approximately the same, while $D\to 0$ for  $K<0.98$, as expected. For the HM the situation is similar but the critical point obtained numerically ($K\approx 0.1$) differs from $K^*\approx 0.2$. 
Thus, for the HM, there is no apparent relation between the maxima observed in the first and second rows of Fig.~\ref{fig:M3quant_class} and diffusion. We should however note that diffusion in the Harper (with $K_1=K_2$) and standard maps is fundamentally different.
\section{Conclusion}
We studied numerically the non-Markovian behaviour of an environment modelled by a quantum kicked map, when it interacts -- pure dephasing --  with a system consisting of a qubit. In particular we centered our attention on the transition from regular to chaotic dynamics. At the extremes, i.e. either regular or  chaotic the behaviour is as expected: if the environment is chaotic then  we expect it to lose memory quicker and be more Markovian than an environment corresponding to regular dynamics. At the transition, where classical dynamics is mixed,  unexpected behaviour manifests in the form of a  peak. In the case of the standard map the peak is located almost exactly at the critical point where the last irrational torus breaks and for dynamics in the cylinder there is unbounded diffusion. For the case of the Harper map, there is no critical point. However we  obtain a peak that is robust to changes in size, time and way of averaging. We conjecture that it should also correspond to a transition point in the classical dynamics. To support this conjecture we studied the fluctuations of the distance between classical trajectories (with no periodic boundaries). We found peaks at locations compatible with the results obtained for the non-Markovianity  measure used.

Additionally, by studying the dependence of no-Markovianity on the initial state of the environment we first found 
that  the main contributions to average non-Markovian behaviour come, not from regular (integrable) 
islands, but from the regions between chaotic and integrable which typically are  complex and composed of broken tori.We were able to build classical phase space pictures from the non-Markovianity measure, where the borders between chaos and regularity are clearly highlighted. 
It is worth remarking that from our numerical investigations yet another feature of quantum fidelity has been unveiled: the long time fluctuations can help identify complex phase space structures like the border between chaotic and regular regions. Traditional (average) fidelity decay approaches have the aim of identifying  sensitivity to perturbations, and chaos. The approach presented here, in contrast, can -- from the fidelity as a function of each individual initial state -- provide a clear image of the classical phase portrait and not just a global quantity from which to infer chaotic (or regular) behavior.
We think that our numerics fit well within the scope of recent experimental setups \cite{Lemos2012}, and some of our findings could be explored.

Finally, we should acknowledge that the validity and interpretation of the
quantities used to asses how non- Markovian a quantum evolution is (including
the BLP measure used here), is a subject of ongoing discussion and which remains
to be decided. Even the ability of a measure of determining whether a system is
Markovian is a topic of intense debate. The BLP measure tied to
the system studied provided new insight particularly in the relation with the
complexities of phase space.  We think that the approach proposed here, i.e. a
simple system coupled to a completely known environment with a feature-rich
classical phase space, could provide with benchmarking possibilities for
non-Markovianity measures.
      
\section*{Acknowledgments}
We thank J. Goold and P. Haikka for stimulating discussions.
 C.P. received support from the  projects
CONACyT 153190 and  UNAM-PAPIIT IA101713 and ``Fondo Institucional del CONACYT''. I.G.M. and D.A.W.
received support from ANCyPT (PICT 2010-1556), UBACyT,
and CONICET (PIP 114-20110100048 and PIP 11220080100728). All three authors are part of a binational grant
(Mincyt-Conacyt \textbf{MX/12/02}).

\appendix
\section{Semiclassical expression for short time decay of the fidelity amplitude}
The fidelity or Loschmidt echo is the quantity originally proposed by Peres \cite{Peres1984} to characterize
sensitivity of a system to perturbation and then used to characterize quantum chaos. It is defined as 
\begin{equation}
M(t)=|f(t)|^2
\end{equation}
where 
\begin{equation}
f(t)=\brac{\psi_0}e^{iH_\epsilon t/\hbar}e^{-i H_0 t /\hbar}\ket{\psi_0}
\end{equation}
where $H_\epsilon$ is differs from $H_0$ by a perturbation term, usually taken as an additive $\epsilon V$ term, with $\epsilon$ a small number.
 
Using the initial value representation for the Van Vleck semiclassical propagator and a concept known as dephasing representation (DR), justified by the shadowing theorem, recently the following simplified semiclassical expression for the fidelity amplitude to \cite{vanicek2003,vanicek2004,vanicek2006} was obtained  
\begin{equation}
\label{DR}
\ODR=\int dq dp W_{\psi}(q,p)\exp(-i\Delta S_\epsilon(q,p,t)/\hbar).
\end{equation}
In \equa{DR} $ W_{\psi}(q,p)$ is the Wigner function of the initial state $\psi$ and 
\begin{equation}
\Delta S_\epsilon(q,p,t)=-\epsilon\int_0^t d\tau V(q(\tau),p(\tau))
\end{equation}
is the action difference evaluated along the \emph{unperturbed} classical trajectory.
 
For sufficiently chaotic system we can approximate the dynamics as random-uncorrelated and  express 
the average fidelity amplitude as \cite{garciamaNJP} 
\begin{equation}
\langle \ODR\rangle = \left[\frac{1}{N}\sum_j \exp(-i\Delta S_{\epsilon,j}/\hbar) \right]
\end{equation}
the average is done over a complete set labeled $j$ ($N$ is the dimension of the  Hilbert space) and $\Delta S_{\epsilon,j}$ is the action difference for the state $j$ at time $t=1$  (we focus on discrete time (maps), so for us it means after one step).
For large enough $N$ we can approximate by a continuous expression
\begin{equation}
\langle \ODR \rangle\approx\left(\int dqdp \exp[-i\Delta S_\epsilon(q,p)/\hbar]\right)^t
\end{equation} 
 
The short time decay of the AFA can be approximated by 
\begin{equation}
|\langle f(t)\rangle|\approx e^{-\Gamma t}
\end{equation}
with 
\begin{equation}
\Gamma\approx-\ln \left|\int dqdp \exp[-i\Delta S_\epsilon(q,p)/\hbar]\right|
\end{equation}
This expression if exact for $t=1$ and is valid for larger times the more chaotic is the system (see \cite{garciamaNJP}).
For both maps in Eqs.~(\ref{standardU}) and (\ref{harperU}) we have $V=K\cos[2\pi q]$ so 
\begin{equation}
\Gamma\approx-\ln\left|\int dq e^{-(\delta K/\hbar)\cos[2 \pi q_i]}\right|= -\ln[|J_0(\delta K/\hbar)|],
\end{equation}
where $J_0$ is the Bessel function of the first kind (with $n=0$), which is an oscillating function.
When $J_0=0$, $\Gamma$ diverges. This means that near these values for short times fidelity decays almost to zero. Nevertheless -- also depending on how chaotic the system is -- after this strong decay, typically there is a large revival \cite{nacho2011,garciamaNJP}.
\section*{References}
%
\providecommand{\newblock}{}

\end{document}